\documentclass[a4paper,conference]{IEEEtran}

\makeatletter
\def\ps@headings{%
\def\@oddhead{\mbox{}\scriptsize\rightmark \hfil \thepage}%
\def\@evenhead{\scriptsize\thepage \hfil \leftmark\mbox{}}%
\def\@oddfoot{}%
\def\@evenfoot{}}
\makeatother 

\IEEEoverridecommandlockouts

\usepackage{caption}
\usepackage{algorithm}
\usepackage{clrscode}

\usepackage{bbm}
\usepackage{longtable}
\usepackage{tabu}
\usepackage{amsfonts}
\usepackage[dvips]{graphicx}
\usepackage{times}
\usepackage{cite}
\usepackage{amsmath}
\usepackage{array}
\usepackage{amssymb}

\usepackage{stfloats}
\usepackage{slashbox}
\usepackage{graphicx}
\usepackage{footnote}
\usepackage{amsthm}
\usepackage{booktabs}
\usepackage{array}
\usepackage{algorithmic}
\usepackage{algorithm}
\usepackage{subeqnarray}
\usepackage{cases}
\usepackage{threeparttable}
\usepackage{color}
\usepackage{epstopdf}
\usepackage{bm}
\usepackage{subcaption}
\usepackage{dsfont}
\usepackage{enumerate}
\usepackage[labelformat=simple]{subcaption}

\usepackage[numbers,sort&compress]{natbib}
\begin{document}

\title{Joint Transceiver Design Based on Dictionary Learning Algorithm for SCMA}
\author{\IEEEauthorblockA{Shanshan Zhang, Wen Chen, and Shaoyuan Chen}
\IEEEauthorblockA{SICS, Department of Electronic Engineering, Shanghai Jiao Tong University, China\\
	Email: $\lbrace$ shansz,wenchen,shaoyuanchen $\rbrace$@sjtu.edu.cn}}
\maketitle

\begin{abstract}
With the explosively increasing demands on the network capacity, throughput and number of connected wireless devices, massive connectivity is an urgent problem for the next generation wireless communications. 
In this paper, we propose a grant-free access protocol for massive connectivity that utilizes a large number of antennas in a base station (BS) and is expected to be widely deployed in cellular networks. The scheme consists of a sparse structure in sparse code multiple access (SCMA) and receiver processing based on dictionary learning (DL). A large number of devices can transmit data without any scheduling process. Unlike existing schemes,  whose signal schedulings require a lot of overhead, the scheduling overhead required by the proposed scheme is negligible, which is attractive for resource utilization and transmission power efficiency. The numerical results show that the proposed scheme has promising performance in massive connectivity scenario of cellular networks. 
\end{abstract}

\begin{IEEEkeywords}
SCMA, dictionary learning (DL), grant-free, massive connectivity, transceiver design
\end{IEEEkeywords}

\section{Introduction}
 Massive machine-type communication (mMTC) or massive Internet of Things (IoT) is one of the key application scenarios of future wireless communication networks. In an mMTC network, hundreds or even thousands of user devices are associated with a single cellular base station (BS), with only a small fraction of them being active. The BS is required to dynamically identify the active users and reliably receive their messages \cite{bigamp_dl}.


To meet the requirements of massive connectivity, sparse code multiple access (SCMA), a nonorthogonal codebook-based multiple access method, was proposed for multiple user access \cite{weifan_5G}. Since the overload feature of SCMA can increase the number of users accessing to the network, it is widely studied for massive connectivity. In the SCMA system, message passing algorithm (MPA) is a conventional algorithm used to decode the transmitted data, which needs to iterate over all the users \cite{7752784}. However, in the massive connectivity scenarios, it is known that the proportion of simultaneously active users in the wireless network normally doesn't exceed 10\% even when the network is busy. Thus, to accelerate the data decoding, it is better for the BS to identify only the active users in the system before decoding the data, rather than iterate over all the users.

In Long Term Evolution (LTE), dynamic user scheduling is achieved through a request grant process. However, the handshake between the BS and the active user will induce a great deal of signaling overhead and system latency. To solve this problem, a signature-based multiple access protocol was studied in \cite{chap4_mtm,chap4_mimo}, where each active user randomly selects a signature sequence (preamble) and sends it to the BS. If the selected preamble is not used by any other user, the active user can establish a connection with the BS. However, contention-based protocols suffer from potential conflicts, and due to the large number of potential users, the contention phase may introduce excessive overhead for control signaling. Therefore, grant-free protocol is more desirable in large-scale device networks, where a user device initiates data transmission without any handshake process with the BS and the overhead required by signal scheduling is negligible. A few of grant-free schemes have been proposed for SCMA \cite{chap1enSCMA2,chap1SCMA3,W3}. In \cite{chap1enSCMA2}, a time-frequency resource called the contention transmission unit (CTU) is defined for uplink grant-free SCMA. A proof-of-concept (PoC) was performed to prove the feasibility and effectiveness of grant-free SCMA in a user-centric cell-free (UCNC) system \cite{chap1SCMA3}. \cite{W3} proposed a message-passing receiver for the uplink grant-free SCMA that performs joint estimation iteratively.

In this paper, we study the massive connectivity of cellular systems with multi-antenna BS. Specifically, we propose a access protocol that enables uplink grant-free transmission of data in SCMA system. The scheme consists of a sparse structure based on SCMA codewords and receiving processing based on dictionary learning (DL). Each active device constructs a data frame based on its own codebook and symbol label. From the received signal, the BS uses DL to recover the transmission frame carrying data symbol and device identity. The sparsity of the framework makes it possible to detect the signal based on DL, while channel estimation and multi-user detection can be achieved simultaneously. Therefore, the proposed scheme reduces transmission overhead and improves decoding performance of the system, which enables low-latency and high-reliability transmission of end devices.

The rest of the paper is organized as follows. Section \uppercase\expandafter{\romannumeral2} introduces the system model and DL. In section \uppercase\expandafter{\romannumeral3}, we design the transmission frame structure and study how to achieve joint user identification and data detection by introducing a DL algorithm named bilinear generalized approximate message
passing (BiG-AMP). In section \uppercase\expandafter{\romannumeral4}, the simulation results are shown to characterize the performance of the proposed method. The conclusion is drawn in section \uppercase\expandafter{\romannumeral5}.

{\it{Notation:}} Throughout, lowercase letters $x$, bold lowercase letters $\textbf{x}$ and bold uppercase letters $\textbf{X}$ denote variables, vectors and matrices, respectively. We use $(\cdot)^*$ to denote complex conjugate. $\textbf{X}^T$ and $\textbf{X}^H$ denote transpose and conjugate transpose of $\textbf{X}$, respectively. We use $x_{n,l}$ or $(\textbf{X})_{n,l}$ to denote the entry of $\textbf{X}$ in the $n$th row and $l$th column. $\mathcal{CN}(x;{{\hat x}},v^{x})$ denotes the complex Guassian probability distribution function for a complex random variable $x$ with mean ${\hat x}$ and variance $v^{x}$. $||\cdot||_2$ and  $||\cdot||_0$ represent 2-norm and 0-norm, respectively.

\section{SCMA System and Dictionary Learning}

\subsection{SCMA System Model}
\label{1.1}

SCMA is proposed with the following properties \cite{3}: i) binary domian data are directly encoded to multidimensional complex domain codewords selected from a predefined codebook set, ii) multiple access is achievable by generating multiple codebooks one for each layer or user, iii) codewords of the codebooks are sparse so that DL algorithms can be used to solve the problems of channel estimation and data detection, iv) the system can be overloaded such that the number of multiplexed layers can be more than spreading factor.

We discuss an uplink grant-free SCMA system with $N$ active users, $K$ subcarriers and $J$ antennas in the BS, where $\mathcal{N}$ represents the active user set. At the transmitting end, $\textbf{c}_{n,i}$, which is the $i$th coded bits of user $n$, is  mapped to codeword $\textbf{x}_{n,i}$ by multi-dimensional SCMA codebooks, where $\textbf{x}_{n,i}=(x_{n,i,1},x_{n,i,2},\ldots,x_{n,i,K})^T$ is a $K$-dimension sparse vector having $d_f$ non-zero elements. At the receiving end, the $i$th symbol of subcarriers $k$ received by antenna $j$ can be expressed as
\begin{eqnarray}
y_{j,i,k}=\sum\limits_{n=1}^N h_{j,n,k}x_{n,i,k} +z_{j,i,k}\label{eq:chap40},
\end{eqnarray}
where $i=1,2,\ldots,I$ and $I$ is the number of symbols in a time slot. $z_{j,i,k}$ represents channel noise added to the $i$th symbol of subcarrier $k$ which follows the Gaussian white noise distribution with $ \mathcal {CN} (0, \delta ^ 2) $, where  $ \delta $ is the standard deviation of $z_{j,i,k}$.

Assume that the channel exhibits an independent and quasi-static flat fading, so it stays the same in each time slot. And also assume that the channel conditions are the same for different subcarriers of a user. Under the above assumptions, we have $h_{j,n,1}=h_{j,n,2}=\ldots=h_{j,n,K}=h_{j,n}$. Then the channel fading model can be established as ${\textbf{h}_n} = \sqrt {{\beta _n}} {\textbf{g}_n}$, where $\textbf{g}_n\backsim \mathcal{CN}(0,\textbf{I}_J)$ is Rayleigh fading, and $ \beta _n $ is the fading coefficient and $\textbf{h}_{n}=(h_{1,n},h_{2,n},\ldots,h_{J,n})^T$. Therefore, (\ref{eq:chap40}) can be rewritten as 
\begin{eqnarray}
y_{j,i,k}=\sum\limits_{n=1}^N h_{j,n}x_{n,i,k} +z_{j,i,k}\label{eq:chap401}.
\end{eqnarray}

For a user $n$, $n\in \mathcal{N}$, the modulated data $\textbf{x}_n\in \mathbb{C}^{L\times 1}$ is 
\begin{eqnarray}
\begin{split}
\textbf{x}_n=&(x_{n,1,1},x_{n,1,2},\ldots,x_{n,1,K},\ldots,\\&x_{n,i,1},\ldots,x_{n,i,K},\ldots,x_{n,N,1},\ldots,x_{n,N,K})^T,
\end{split}
\label{eq:chap403}
\end{eqnarray} 
 where $L=K \times I$.
  For all users, we assume a power constraint $P$, i.e.,
\begin{eqnarray}
\frac{1}{L}E[\textbf{x}_n^H\textbf{x}_n] \le P,\forall n \in \mathcal{N}. \label{eq:chap41}
\end{eqnarray}
So the received signal can be expressed as
\begin{eqnarray}
\textbf{Y} = \sum\limits_{n \in \mathcal{N}} {{\textbf{h}_n}{\textbf{x}^{T}_n}}  + \textbf{Z}, \label{eq:chap43}
\end{eqnarray}
where  $\textbf{Y}=(\textbf{y}_{1},\textbf{y}_{2},\ldots,\textbf{y}_{J})^T\in \mathbb{C}^{J\times L}$,  $\textbf{Z}=(\textbf{z}_{1},\textbf{z}_{2},\ldots,\textbf{z}_{J})^T\in \mathbb{C}^{J\times L}$ and 
\begin{eqnarray}
\begin{split}
\textbf{y}_j=&(y_{j,1,1},y_{j,1,2},\ldots,y_{j,1,K},\ldots,\\&y_{j,i,1},\ldots,y_{j,i,K},\ldots,y_{j,N,1},\ldots,y_{j,N,K})^T,
\end{split}\\
\label{eq:chap404}
\begin{split}
\textbf{z}_j=&(z_{j,1,1},z_{j,1,2},\ldots,z_{j,1,K},\ldots,\\&z_{j,i,1},\ldots,z_{j,i,K},\ldots,z_{j,N,1},\ldots,z_{j,N,K})^T.
\end{split}
\label{eq:chap405}
\end{eqnarray} 
Then (\ref{eq:chap43}) can be rewritten as
\begin{eqnarray}
\textbf{Y} = \textbf{H}\textbf{X} +\textbf{Z},
\label{eq:chap451}
\end{eqnarray}
where $\textbf{H}=(\textbf{h}_1,\textbf{h}_2,\ldots,\textbf{h}_N)\in \mathbb{C}^{J\times N}$ and $\textbf{X}=(\textbf{x}_1,\textbf{x}_2,\ldots,\textbf{x}_N)^T\in \mathbb{C}^{N\times L}$ are channel transmission matrix and signal matrix, respectively.

\subsection{Dictionary Learning}
\label{sec:bigamp}

Firstly, we briefly introduce DL, which is a key component of the proposed scheme. Sparse signal processing, especially compressed sensing, has attracted widespread attention in the field of signal processing and wireless communication \cite{7842611}. Compressed sensing or sparse recovery refers to a type of signal processing technique that recovers sparse vectors from incomplete linear measurements \cite{4472240}. As a kind of compressed sensing technique, DL aims to learn the dictionary matrix $ \textbf {H} $ from the observed signal. In other words, for the model
\begin{eqnarray}
\textbf{y}_l = \textbf{H}\textbf{x}_l +\textbf{z}_l,l=1,2,3,\ldots,L,
\label{eq:chap4511}
\end{eqnarray}
where $\textbf{x}_l$ is a sparse vector,  $\textbf{y}_l$ is an observed data vector and $\textbf{z}_l$ is a noise vector, DL aims to find a dictionary matrix $\hat{\textbf{H}}$ and  a sparse vector $\hat{\textbf{x}}_l$ which satisfy $\textbf{y}_l \approx \hat{\textbf{H}}\hat{\textbf{x}}_l$. The joint estimation of dictionary matrix and sparse vector is a highly underdetermined bilinear problem when there is no constraint. Taking the sparsity hypothesis into account, we can restate the problem by finding the most sparse solution, i.e.,
\begin{eqnarray}
\arg\min_{\textbf{H},\{\textbf{x}_l\}}&\sum_{l=1}^L{\lVert \textbf{y}_l-\textbf{H}\textbf{x}_l\rVert}_{2}^2,\\ 
s.t.&{\lVert \textbf{x}_l\rVert}_{0}\le T_0,
\label{eq:chap4512}
\end{eqnarray}
where $T_0$ is the maximum sparsity contraint. Although it is not easy to solve this problem, various algorithms have been developed on this optimization problem. Experience implies that if $\textbf{x}_l$ is sparse enough and a large amount of data is available,  $\textbf{x}_l$ and $\textbf{H}$ can be recovered accurately \cite{chap4_bigamp2}. Although algorithms differ with respect to the dictionary updating strategy, they  all seek solutions through the block coordinate descent process, where the dictionary matrix and the sparse vector are updated alternately.

The DL problem can be viewed as the decomposition of a matrix $\textbf{Y}$ of (\ref{eq:chap451}),
where $\textbf{Y} = (\textbf{y}_1,\textbf{y}_2,\ldots,\textbf{y}_L)\in \mathbb{C}^{J\times L}$, $\textbf{X} = (\textbf{x}_1,\textbf{x}_2,\ldots,\textbf{x}_L)\in \mathbb{C}^{N\times L}$ and $\textbf{Z} = (\textbf{z}_1,\textbf{z}_2,\ldots,\textbf{z}_L)$. It aims to decompose $\textbf{Y}$ into the product of $\hat{\textbf{H}}$ and $\hat{\textbf{X}}$. It is worth noting that the solution to the DL problem has two ambiguities. The first one is phase ambiguity. Defining $\bar{\textbf{H}}=\hat{\textbf{H}}\mathbf{\Gamma}$ and $\bar{\textbf{X}}={\mathbf{\Gamma}}^{-1} \hat{\textbf{X}}$, where $\mathbf{\Gamma}$ is a diagonal matrix, it is easy to see that $\hat{\textbf{H}}\hat{\textbf{X}}=\bar{\textbf{H}}\bar{\textbf{X}}$, i.e., if $\hat{\textbf{H}}$ is a solution to the DL problem, another solution $\bar{\textbf{H}}$ can be obtained by scaling any column. At this time, $\bar{\textbf{X}}$ scales $\hat{\textbf{X}}$ in lines. The second one is permutation ambiguity, since the permutation of one solution could produce another solution. Suppose $\mathbf{\Pi}$ is an $N\times N$ permutation matrix, then $\tilde{\textbf{H}}=\hat{\textbf{H}}\mathbf{\Pi}$ and $\tilde{\textbf{X}}={\mathbf{\Pi}}^{-1} \hat{\textbf{X}}$ is obtained by transforming columns of $\hat{\textbf{H}}$ and rows of $\hat{\textbf{X}}$, respectively. Obviously, $\tilde{\textbf{H}}\tilde{\textbf{X}}=\hat{\textbf{H}}\hat{\textbf{X}}$, and the sparsity of this solution is same as the original solution. Therefore, any DL algorithm that finds the most sparse solution cannot solve this two ambiguities. Note that these ambiguities are inherent in the problem and do not depend on the DL algorithm used. Therefore,
they should be handled carefully when applying DL algorithm. In section \uppercase\expandafter{\romannumeral3}, we will show how to use DL for grant-free access in massive connectivity scenarios, and how to solve these ambiguities to achieve user identification, channel estimation and data detection.

\section{Grant-free Access Based on Dictionary Learning}

\subsection{Transmission Frame Structure}

\begin{figure}[h]
	\begin{center}
		\includegraphics[scale=.35]{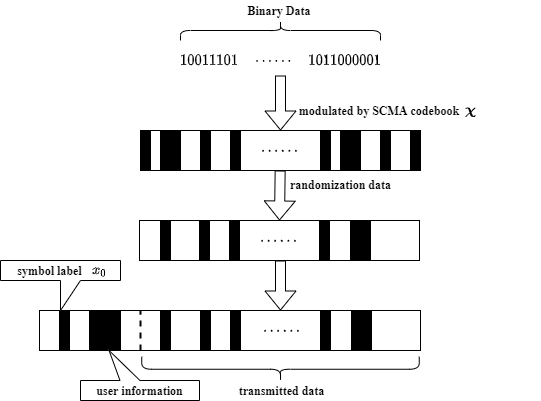}
	\end{center}
	\vspace{-0.3em}
	\caption{\small{Transmitter data processing.}}
	\label{fig:chap41}
	\vspace{-0.2em}
\end{figure}

In view of the sparsity of the SCMA codebooks, the design allows zero symbols to be randomly and independently distributed in the data encoded by the SCMA encoder. In particular, represent $ \mathcal {C} $ as the SCMA-encoded complex constellation point set, where $ 0 \notin \mathcal {C} $, and the probability distribution of each symbol that having independent and identical distribution is as follows
\begin{eqnarray}
P(x)  = \left\{ \begin{array}{l}
1 - \gamma ,x = 0\\
\frac{\gamma }{{\left| \mathcal{C}  \right|}},x \in \mathcal{C} 
\end{array} \right.,
\label{eq:chap44}
\end{eqnarray}
where $ \gamma = d_f / K \in (0,1) $ represents the level of sparsity, and $ |\mathcal{C}| $ means the number of element in $ \mathcal{C} $. Then a priori probability of $x_{n,l}$ can be expressed as (\ref{eq:chap416}) by randomizing the position of non-zero values in the SCMA codebooks, 
\begin{eqnarray}
p_{x_{n,l}}(x_{n,l}) =(1 - \gamma )\delta (x_{n,l}) + \frac{\gamma }{{\left| \mathcal{C}  \right|}}\delta (x_{n,l} - x), x \in \mathcal{C},
\label{eq:chap416}
\end{eqnarray}
where $\delta(\cdot)$ represents Dirac delta function. 

The encoding process of the SCMA system is shown in Fig. \ref{fig:chap41}. $ \mathcal {X} $ represents the SCMA system user codebooks. {\it{User information}} is the information for identifying user and solving permutation ambiguity at the receiving end. The symbol label $ x_0 $ is known to both the transmitting and receiving ends and is set for correcting the phase ambiguity. This work will be introduced in section \ref{B}. It needs to note that $L=K\times I+K+1$ with $K$ user information symbol and one symbol label in the following.

\subsection{Dictionary Learning Based on BiG-AMP Algorithm}
\label{A}

Recalling (\ref{eq:chap451}), we expect that the channel estimation and data detection can be acheived by decomposing $\textbf{Y}$ into $\textbf{H}$ and $\textbf{X}$ without ignoring the influence of noise. To this end, we first rewrite (\ref{eq:chap451}) as the form of (\ref{eq:chap432}), i.e.,
\begin{eqnarray}
\textbf{y}_l = \textbf{H}\textbf{x}_l +\textbf{z}_l,
\label{eq:chap432}
\end{eqnarray}
where $\textbf{y}_l$ and $\textbf{x}_l$ are the $l$th column of $\textbf{Y}$ and $\textbf{X}$, respectively. Recalling that in section \ref{1.1} $\textbf{X}$ is sparse, thus $\textbf{x}_l$ is sparse and $\textbf{y}_l$ is sparse linear combination of columns of $\textbf{H}$. Then the joint estimation of channel transmission matrix and sparse data can be transformed into a DL problem, i.e.,
\begin{eqnarray}
(\hat{\textbf{H}},\hat{\textbf{X}})=\arg\min_{\textbf{H},\textbf{X}}\sum\limits_{l=1}^L{\lVert \textbf{x}_l\rVert}_{0},\\ 
s.t.
\sum\limits_{l=1}^L{\lVert \textbf{y}_l-\textbf{H}\textbf{x}_l\rVert}_{2}^2\le \epsilon.
\label{eq:chap433}
\end{eqnarray}

A variety of DL algorithms have been developed and empirically proven to have good performance under sparse and massive connectivity conditions \cite{bigamp_dl}. In the following, a DL algorithm is employed to decompose the received signal matrix into a dictionary matrix and a sparse matrix.

\begin{table}[htbp]
	\centering  
	\label{tab:chap41}  
	\resizebox{\linewidth}{!}{
		\begin{tabular}{l r}  
			\hline  
			\textbf{Algorithm 1:} BiG-AMP algorithm&\\
			\hline
			\textbf{Input:} $\boldsymbol{Y},p_{h_{j,n}} ,p_{x_{n,l}}$& \\
			\textbf{Initialization:}$\forall j,n,l,$ generating ${h_{j,n}}$randomly from $p_{h_{j,n}}$,$v_{j,n}^h\left( 1 \right) = \bar \beta ,$\\$\;{\hat x_{n,l}}\left( 1 \right) = {E_x},v_{n,l}^x\left( 1 \right) = {\rm{\sigma }}_x^2,and\;{\hat s_{j,l}}\left( 0 \right) = 0.$&\quad \\
			
		 for \quad $t=1,2,\cdots ,T_{max}$ (iteration) & \quad \\
			 \quad $\forall j,n:\bar v_{j,l}^w\left( t \right) = \mathop \sum \limits_{n = 1}^N [|{\hat h_{j,n}}\left( t \right){|^2}v_{n,l}^x\left( t \right) + |{\hat x_{n,l}}\left( t \right){|^2}v_{j,n}^h\left( t \right)]$ &R1\\
			\quad $\forall j,l:{\bar w_{j,l}}\left( t \right) = \mathop \sum \limits_{n = 1}^N {\hat h_{j,n}}\left( t \right){\hat x_{n,l}}\left( t \right)$ &R2\\
			\quad 
			$\forall j,l:v_{j,l}^w\left( t \right) = \bar v_{j,l}^w\left( t \right) + \mathop \sum \limits_{n = 1}^N v_{n,l}^x\left( t \right)v_{j,n}^h\left( t \right)$&R3\\
			\quad 
			$\forall j,l:{\hat w_{j,l}}\left( t \right) = {\bar w_{j,l}}\left( t \right) - {\hat s_{j,l}}\left( {t - 1} \right)\bar{v}_{j,l}^w\left( t \right)$&R4\\
			\quad 
			$\forall j,l:v_{j,l}^z\left( t \right) = \bar v_{j,l}^w\left( t \right){\sigma ^2}{[\bar v_{j,l}^w\left( t \right) + {\sigma ^2}]^{ - 1}}$&R5\\
			\quad 
			$\forall j,l:{\hat z_{j,l}}\left( t \right) = \bar v_{j,l}^w\left( t \right)\left[ {{y_{j,l}}\left( t \right) - {{\hat w}_{j,l}}\left( t \right)} \right]{[\bar v_{j,l}^w\left( t \right) + {\sigma ^2}]^{ - 1}} + {\hat w_{j,l}}\left( t \right)$&R6\\
			\quad 
			$\forall j,l:v_{j,l}^s\left( t \right) = \left[ {1 - v_{j,l}^z\left( t \right)/v_{j,l}^w\left( t \right)} \right]/v_{j,l}^w\left( t \right)$&R7\\
			\quad 
			$\forall j,l:{\hat s_{j,l}}\left( t \right) = [{\hat z_{j,l}}\left( t \right) - {\hat w_{j,l}}\left( t \right)]/\bar v_{j,l}^w\left( t \right)$&R8\\
			\quad 
			$\forall j,n:v_{j,n}^q\left( t \right) = {[\mathop \sum \limits_{l = 1}^L |{\hat x_{n,l}}\left( t \right){|^2}v_{j,n}^s\left( t \right)]^{ - 1}}$&R9\\
			\quad 
			$\forall j,n:{\hat q_{j,n}}\left( t \right) = {\hat h_{j,n}}\left( t \right)\left[ {1 - v_{j,n}^q\left( t \right)\mathop \sum \limits_{l = 1}^L v_{n,l}^x\left( t \right)v_{j,l}^s\left( t \right)} \right] + v_{j,n}^q\left( t \right)\mathop \sum \limits_{l = 1}^L \hat x_{n,l}^*\left( t \right){\hat s_{j,l}}\left( t \right)$&R10\\
			\quad 
			$\forall n,l:v_{n,l}^r\left( t \right) = {[\mathop \sum \limits_{j = 1}^J |{\hat h_{j,n}}\left( t \right){|^2}v_{j,l}^s\left( t \right)]^{ - 1}}$&R11\\
			\quad 
			$\forall n,l:{\hat r_{n,l}}\left( t \right) = {\hat x_{n,l}}\left( t \right)\left[ {1 - v_{n,l}^r\left( t \right)\mathop \sum \limits_{j = 1}^J v_{j,n}^h\left( t \right)v_{j,l}^s\left( t \right)} \right] + v_{n,l}^r\left( t \right)\mathop \sum \limits_{j = 1}^J \hat h_{j,n}^*\left( t \right){\hat s_{j,l}}\left( t \right)$&R12\\
			\quad
			$\forall j,n:{\hat h_{j,n}}\left( {t + 1} \right) = {E_{{p_{{h_{j,n}}|Y}}}}[{h_{j,n}}|{\hat q_{j,n}}\left( t \right),v_{j,n}^q\left( t \right)]$&R13\\
			\quad 
			$\forall j,n:v_{j,n}^h\left( {t + 1} \right) = {E_{{p_{{h_{j,n}}|Y}}}}[{\left| {{h_{j,n}} - {{\hat h}_{j,n}}\left( {t + 1} \right)} \right|^2}|{\hat q_{j,n}}\left( t \right),v_{j,n}^q\left( t \right)]$&R14\\
			\quad 
			$\forall n,l:{\hat x_{n,l}}\left( {t + 1} \right) = {E_{{p_{{x_{n,l}}|Y}}}}[{x_{n,l}}|{\hat r_{n,l}}\left( t \right),v_{n,l}^r\left( t \right)]$&R15\\
			\quad 
			$\forall n,l:v_{n,l}^x\left( {t + 1} \right) = {E_{{p_{{x_{n,l}}|Y}}}}[{\left| {{x_{n,l}} - {{\hat x}_{n,l}}\left( {t + 1} \right)} \right|^2}|{\hat r_{n,l}}\left( t \right),v_{n,l}^r\left( t \right)]$&R16\\
			\quad \textbf{If}\quad 
			$\mathop \sum \limits_{j = 1}^J \mathop \sum \limits_{l = 1}^L |{\bar w_{j,l}}\left( t \right) - {\bar w_{j,l}}\left( {t + 1} \right){|^2} < {\tau}_{BiG-AMP} \mathop \sum \limits_{j = 1}^J \mathop \sum \limits_{l = 1}^L |{\bar w_{j,l}}\left( t \right){|^2}$,\textbf{stop}&\quad \\
			 \textbf{End}&\quad \\

			\textbf{Output:}
			$\forall j,n:{\hat p_{{h_{j,n}}}}\left( {{h_{j,n}}} \right) \propto {p_{{h_{j,n}}}}\left( {{h_{j,n}}} \right){\cal CN}\left( {{h_{j,n}};{{\hat q}_{j,n}}\left( t \right),v_{j,n}^q\left( t \right)} \right)$&R17 \\
			\quad\quad \quad\quad
			$\forall n,l:\;{\hat p_{{x_{n,l}}}}\left( {{x_{n,l}}} \right) \propto {p_{{x_{n,l}}}}\left( {{x_{n,l}}} \right){\cal CN}\left( {{x_{n,l}};{{\hat r}_{n,l}}\left( t \right),v_{n,l}^r\left( t \right)} \right)$&R18 \\
			\hline
		\end{tabular}
	}
\end{table}

\begin{figure}[t]
	\begin{center}
		\includegraphics[scale=.35]{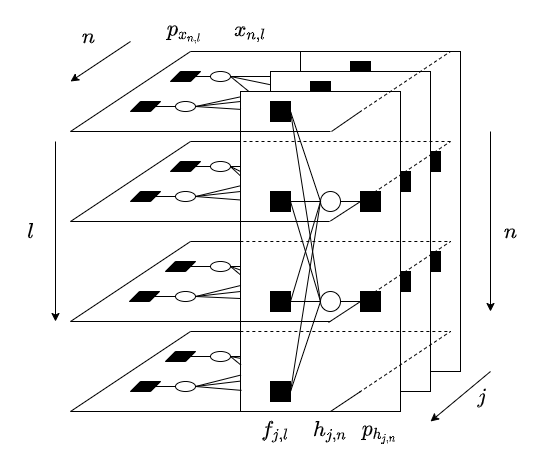}
	\end{center}
	\vspace{-0.3em}
	\caption{\small{Factor graph.}}
	\label{fig:chap42}
	\vspace{-0.2em}
\end{figure}

All kinds of the DL algorithms can be applied to our system, among which the BiG-AMP algorithm is adopted in our study as it makes better use of the prior information of channel transmission matrix and signal matrix, and has been verified the reliability of the performance in various applications \cite{7891590,8063440}. The numerical results show that the DL algorithm based on the BiG-AMP algorithm works well in the massive connectivity scenarios.

Use (\ref{eq:chap48}) to represent the received signal, which is consistent with (\ref{eq:chap432})
\begin{eqnarray}
{y_{j,l}} = \sum\limits_{n = 1}^N {{h_{j,n}}{x_{n,l}}}  + {z_{j,l}},\forall j,l.
\label{eq:chap48}
\end{eqnarray}
From (\ref{eq:chap48}) it can be seen that the received signal $ y_ {j, l} $ is bilinear for its constraint factors $ {h_ {j, n}}$ and $ {x_ {n, l}} $. Eq. (\ref{eq:chap48}) can be represented with a factor graph, as depicted in Fig. \ref{fig:chap42}, where $ {h_ {j, n}}$ and $ {x_ {n, l}} $ are variable nodes that appear as white circles, and $ f_ {j, l} $ is factor node that appears as black square.

In \cite{chap4_bigamp1,chap4_bigamp2}, a BiG-AMP algorithm is proposed to solve such DL problems. The central idea is to approximate the marginal probability distribution functions (PDFs) of $h_{j,n}$ and $x_{n,l}$ through the central limit theorem and the Taylor series. Then the marginal PDFs ${p_{{h_{j,n}}|Y}}({h_{j,n}}|\boldsymbol{Y})$ and ${p_{{x_{n,l}}|Y}}({x_{n,l}}|\boldsymbol{Y}) $ are obtained by (\ref{eq:chap414}) and (\ref{eq:chap415}), respectively, i.e.,
\begin{eqnarray}
{p_{{h_{j,n}}|Y}}({h_{j,n}}|\boldsymbol{Y}) = \frac{{{p_{{h_{j,n}}}}({h_{j,n}})\mathcal{CN}({h_{j,n}};{{\hat q}_{j,n}},v_{j,n}^{(q)})}}{{\int {{p_{{h_{j,n}}}}({h_{j,n}})\mathcal{CN}({h_{j,n}};{{\hat q}_{j,n}},v_{j,n}^{(q)})d{h_{j,n}}} }},
\label{eq:chap414}
\end{eqnarray}
and
\begin{eqnarray}
{p_{{x_{n,l}}|Y}}({x_{n,l}}|\boldsymbol{Y}) = \frac{{{p_{{x_{n,l}}}}({x_{n,l}})\mathcal{CN}({x_{n,l}};{{\hat r}_{n,l}},v_{n,l}^{(r)})}}{{\int {{p_{{x_{n,l}}}}({x_{n,l}})\mathcal{CN}({x_{n,l}};{{\hat r}_{n,l}},v_{n,l}^{(r)})d{x_{n,l}}} }},
\label{eq:chap415}
\end{eqnarray}
where variables $h_{j,n}$ and $x_{n,l}$ follow the Gaussian distributions with means ${\hat q}_{j,n}$ and ${\hat r}_{n,l}$, respectively, and variances $v_{j,n}^{(q)}$ and $ v_{n,l}^{(r)}$, respectively.
These four parameters, i.e., ${\hat q}_{j,n}, {\hat r}_{n,l}, v_{j,n}^{(q)}$ and $ v_{n,l}^{(r)}$ are continuously updated with the iterations. Algorithm 1 summarizes the BiG-AMP algorithm. The algorithm includes a maximum number of iterations $T_{max}$ and a stopping condition based on the residual and a defined parameter ${\tau}_{BiG-AMP}$. Algorithm 1 is briefly described below.

In the initialization phase, the prior probabilities $p_{h_{j,n}}(h_{j,n})$ and $p_{x_{n,l}}(x_{n,l})$ are used to calculate the means and variances of $h_{j,n}$ and $x_{n,l}$. $p_{x_{n,l}}(x_{n,l})$ is described by (\ref{eq:chap416}) and $h_{j,n}\sim \mathcal{CN}(0,\bar \beta )$, where $\bar \beta$ is the average path loss of all users in the SCMA system. In (R1-R2) of Algorithm 1, the means $\bar w_{j,l}(t)$ and variances $\bar v_{j,l}^w(t)$ of $(\textbf{HX})_{j,l}$ are obtained by accumulating the data passed from the variable nodes $h_{j,n}$ and $x_{n,l}$ to  the factor node $f_{j,l}$. In (R3-R4), the adjusted means $\hat w_{j,l}(t)$ and variances $v_{j,l}^w(t)$ of the output data of factor node $f_{j,l}$ are computed  using Onsager correction \cite{CS}. In (R5-R12), the Onsager are further applied to correct the means and variances of variable nodes $h_{j,n}$ and $x_{n,l}$ to obtain messages, which are transmitted from the factor node $f_{j,l}$ to the variable nodes $h_{j,n}$ and $x_{n,l}$. Specifically, (R9) and (R10) calculate means $\hat q_{j,n}\left(t\right)$ and corresponding variances $v_{j,n}^q\left( t \right)$ for each $h_{j,n}$, while (R11) and (R12) compute means $\hat r_{n,l}(t)$ and variances $v_{n,l}^r\left( t \right)$ for each $x_{n,l}$. In (R13-R14), $\hat q_{j,n}\left(t\right)$ and $v_{j,n}^q\left( t \right)$  together generate the posterior mean $\hat h_{j,n}$ and variance $v_{j,n}^h$ by merging with the prior distribution $p_{h_{j,n}}(h_{j,n})$ through (\ref{eq:chap414}). A similar process is applied to each $x_{n,l}$ in (R15-R16), and the posterior mean $\hat x_{n,l}$ and variance $v_{n,l}^x$ can be computed.  Finally, in (R17-R18), the BiG-AMP algorithm outputs the posterior probability estimates $\hat p_{{h_{j,n}}}(h_{j,n})$ and $\hat p_{{x_{n,l}}}(x_{n,l})$. During the calculation process, an adaptive damping is also applied to keep the convergence of the BiG-AMP algorithm \cite{chap4_bigamp1}.

\subsection{Joint User Identification and Data Detection}
\label{B}
In section \ref{A}, we have discussed how to get $({\hat {\bf{H}},\hat {\bf{X}}})$ by BiG-AMP algorithm. However, the estimated result faces the problem of the permutation and phase ambiguities (see Sec. \ref{sec:bigamp}). Since the received signal carries the user information, the influence caused by the permutation matrix can be ignored, and the phase ambiguity can be corrected by using symbol label $ x_0 $.

we first get $\hat{p}_{x_{n,l}}(x_{n,l})$ by the BiG-AMP algorithm, and then the expectation ${E_{{x_{n,l}}}}[{x_{n,l}}]$ can be obtained by integration. Through ${E_{{x_{n,l}}}}[{x_{n,l}}]$, $\hat{x}_{n,l}^{(s)}$ can be obtained as
\begin{eqnarray}
\hat x_{n,l}^{(s)}{\rm{ = }}\left\{ \begin{array}{l}
{E_{{x_{n, l}}}}[{x_{n,l}}],\quad\left| {{E_{{x_{n,l}}}}[{x_{n,l}}]} \right| \ge \tau\\
0,\quad\quad\quad\quad\quad  otherwise
\end{array} \right..
\label{eq:chap410}
\end{eqnarray}

It is implied from (\ref{eq:chap410}) that ${E_{{x_{n,t}}}}[{x_{n,t}}]$ is output when $|{E_{{x_{n,t}}}}[{x_{n,t}}]|$ is greater or equal than a threshold $ \tau $, otherwise it is regarded as the inserted symbol zero. Next, we propose a method to recover the signal from phase ambiguity.

For the output signal $ \hat x_{n,l}^{(s)}$, let $\hat{x}_{n,0}^{(s)}$ represents the output of its first non-zero value, i.e., $\hat{x}_{n,0}^{(s)}$ corresponds to the symbol label $x_0$ at the transmitter. Then the phase offset of the $n$th user $\varphi_n$ can be calculated as
\begin{eqnarray}
\varphi_n  = \frac{{{x_0}}}{{\hat x_{n,0}^{(s)}}}.
\label{eq:chap411}
\end{eqnarray}
The output signal without phase ambiguity for user $n$ can be described as
\begin{eqnarray}
\hat x_{n,l}^{(c)}=\varphi_n  \hat x_{n,l}^{(s)}.
\label{eq:chap412}
\end{eqnarray}

Then by matching the user codebook $ \mathcal {X} $ with the user information, the modulation symbol $ \mathcal {C} $ can be obtained, i.e., the modulated data ${\hat{ \textbf{x}}_n} = ({\hat x_{n,1}},{\hat x_{n,2}}, \cdots ,{\hat x_{n,L}})$, $n = 1,2, \cdots ,N$, can be obtained as depicted by (\ref{eq:chap413}). Finally, by demodulating $\hat{ \textbf{x}}_n$, the binary data $\hat{ \textbf{c}}_n$ can be recovered through the user codebook.

\begin{eqnarray}
\hat x_{n,l}= \left\{ \begin{array}{l}
\mathop {\arg \min }\limits_{x \in \mathcal{C} } {\left| {\hat x_{n,l}^{(c)} - x} \right|^2},\quad \hat x_{n,l}^{(c)} \ne 0\\
0,\quad\quad\quad\quad\quad\quad\quad\quad otherwise
\end{array} \right.
\label{eq:chap413}
\end{eqnarray}

\section{Simulation Results}

Since BiG-AMP is a bilinear estimation algorithm based on a large amount of data, the classical model with four subcarriers and six users is no longer suitable for the SCMA system simulation. Therefore, in the simulation of the SCMA system based on BiG-AMP algorithm, the numbers of subcarriers and users should be appropriate and related to the sparse level $ \gamma $. In the simulation, the number of valid subcarriers $ d_f = 2 $ is selected, and the number of users carried by each subcarrier is $ d_v = 3 $, i.e., the overload rate $ \lambda=d_v/d_f=1.5 $ remains the same as the classical model. Thus, the user codebook can be set on the original constellation point. The size of codebook is $M$, and $ M = | \mathcal{X} | = 4 $. For the case where $ d_f$ and $ d_v $ are determined, there comes the number of subcarriers $ K = d_f / \gamma $, and the number of users $ N = d_v / \gamma $. The number of symbols $I$ in a time slot is set as 1000. Finally, the number of antennas, under the requirements of the BiG-AMP with $ J \ge N $, is designed to be $ J = N $.

\subsection{Performance of the BiG-AMP algorithm used in the SCMA system when the  sparse level $ \gamma $ is different}

We investigate the performance of the BiG-AMP algorithm applied to the SCMA system with different sparse levels  $\gamma$, i.e.,

\begin{itemize}
	\item $\gamma=0.25,K=8,N=12$;
	\item  $\gamma=0.20,K=10,N=15$;
	\item  $\gamma=0.10,K=20,N=30$.
\end{itemize}

\begin{figure}[h]
	\begin{center}
		\includegraphics[scale=.5]{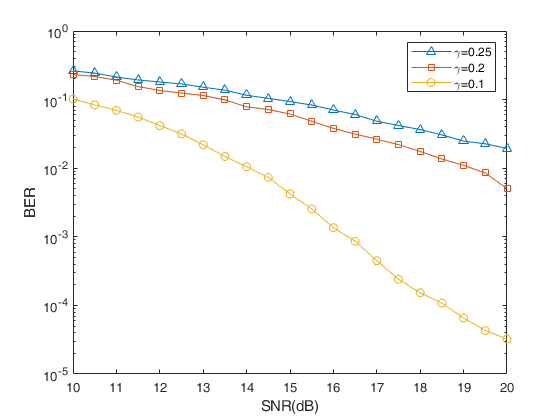}
	\end{center}
	\vspace{-0.3em}
	\caption{\small{BER curves based on BiG-AMP with different $\gamma$.}}
	\label{fig:chap44}
	\vspace{-0.2em}
\end{figure}

Fig. \ref{fig:chap44} shows the bit error rates (BERs) versus signal-to-noise ratio (SNR) based on BiG-AMP with different sparse levels. It can be observed that the more sparse the data, the better the BER performance for user codebooks with different sparse levels. For example,
 when the SNR is 17.5dB, the BER for $ \gamma = 0.25 $ is about $0.04$, whereas the BER for $ \gamma = 0.2 $ is about $0.02$, and for $ \gamma = 0.1 $ the BER can be as low as about $ 2\times 10 ^ {-4} $. Obviously, the sparse level of data has an important impact on BiG-AMP algorithm.

\subsection{Performance comparison of BiG-AMP and NP-LSD-MPA algorithms}

\begin{figure}[h]
	\begin{center}
		\includegraphics[scale=.5]{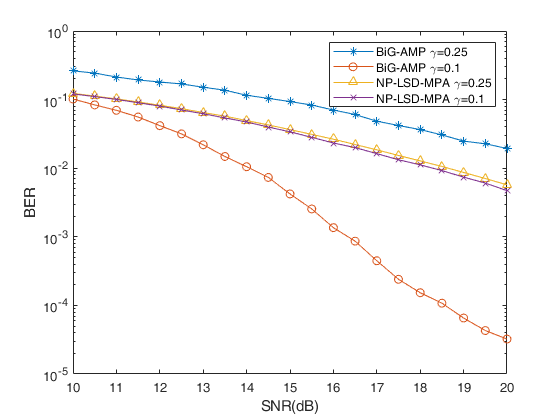}
	\end{center}
	\vspace{-0.3em}
	\caption{\small{BER curves of BiG-AMP and NP-LSD-MPA algorithms.}}
	\label{fig:chap45}
	\vspace{-0.2em}
\end{figure}

List sphere decoding based  message passing algorithm with
node prunning (NP-LSD-MPA) \cite{7752784} is an excellent decoding algorithm for SCMA. The performance comparison between BiG-AMP and NP-LSD-MPA algorithms is shown in Fig. \ref {fig:chap45}. The simulation result is obtained under the condition of sparse levels $ \gamma = 0.1 $ and $ \gamma = 0.25 $. It can been seen that NP-LSD-MPA has better performance for $ \gamma = 0.25 $ and BiG-AMP performs significantly better than NP-LSD-MPA for $ \gamma = 0.1 $. To achieve the same BER level under $ \gamma = 0.1 $, the SNR required by the BiG-AMP algorithm is much lower than that required by the NP-LSD-MPA algorithm. And as the SNR increases, the advantage of the BiG-AMP algorithm becomes more obvious. For example, it can be seen that when the SNR is 14dB, the BER for BiG-AMP is about $0.01$, but for NP-LSD-AMP the BER is about $0.05$. In addition, when the SNR increases to 18.5dB, the BER for NP-LSD-AMP merely reduces to about $10 ^ {-2}$, whereas for BiG-AMP the BER almost reduces to about $10 ^ {-4}$. The result also shows that sparsity has almost no effect on NP-LSD-MPA. Since BiG-AMP algorithm makes full use of the sparsity of transmitted data, it has a better performance when $ \gamma $ is small. Consequently, for the SCMA system, the BiG-AMP algorithm has promising performance for the data detection at the receiving end.

\section{Conclusion}
For the future communication networks, massive connectivity is an emerging research topic. SCMA is a promising non-orthogonal multiple access technique for massive connectivity. We study the uplink grant-free SCMA system with the BiG-AMP algorithm to estimate the channel information and the transmitted data. The proposed method takes advantage of the sparsity of the SCMA codebooks and has promising performance in data detection, which is supposed to have great application value in the future communication environments. In our future work, we will study how to optimize the method continuously.

\section*{Acknowledgement}

This work is supported in part by National Key Project
2018YFB1801102, in part by STCSM 20JC1416502, and in
part by NSFC 61671294 and 62071296.

\bibliographystyle{IEEEtran}
\bibliography{IEEEabrv}

\end{document}